\documentclass[12pt,preprint]{aastex}

\newcommand\msun{M_{\odot}}
\newcommand\lsun{L_{\odot}}
\newcommand\rsun{R_{\odot}}
\newcommand\msunyr{M_{\odot}\,\rm yr^{-1}}
\newcommand\be{\begin{equation}}
\newcommand\en{\end{equation}}

\newcommand\mdot{\dot{M}}

\newcommand\ha{H$\alpha \;$}

\begin{document}

\shortauthors{Muzerolle et al.}
\shorttitle{Accretion in Young Substellar Objects}

\title{Measuring Accretion in Young Substellar Objects: 
Approaching the Planetary Mass Regime \altaffilmark{1}}

\author{
James Muzerolle\altaffilmark{2},
Kevin L. Luhman\altaffilmark{3},
C\'esar Brice\~no\altaffilmark{4},
Lee Hartmann\altaffilmark{3},
Nuria Calvet\altaffilmark{3}}

\altaffiltext{1}{Based on observations performed at Las Campanas Observatory.
This publication makes use of data products from the Two Micron All
Sky Survey.}
\altaffiltext{2}{Steward Observatory, 933 N. Cherry Ave., 
The University of Arizona, Tucson, AZ 85721}
\altaffiltext{3}{Harvard-Smithsonian Center for Astrophysics, 60
Garden St., Cambridge, MA 02138}
\altaffiltext{4}{Centro de Investigaciones de Astronom{\'\i}a (CIDA),\\
Apartado Postal 264, M\'erida 5101-A, Venezuela}

\begin{abstract}

We present observations of \ha emission line profiles taken at Magellan 
Observatory for a sample of 39 young low-mass stars and brown dwarfs
in the Taurus and Chamaeleon~I star forming regions. 
We have identified 11 new substellar accretors, more than tripling
the number of known brown dwarfs with measurable accretion activity.
These include the lowest-mass objects yet seen with accretion,
with masses down to $\sim0.015$~$\msun$.
Using models of \ha emission produced in magnetospheric accretion flows,
the most widely applicable primary calibrator now available,
we determine the first estimates of mass accretion rates for objects at
such extremely low masses.
For the six objects with masses $\lesssim 0.03 \; \msun$, we find accretion
rates of $\sim 5 \times 10^{-12} \; \msunyr$, among the smallest yet measured.
These new results continue the trend of decreasing mass accretion rate
with decreasing (sub)stellar mass that we have noted previously for samples
of more massive objects; the overall correlation is $\mdot \propto M^{2.1}$,
and now extends over a mass range of over two orders of magnitude. 
Finally, the absence of a discontinuity in the distribution of accretion rates 
with mass tends to suggest that stars and brown dwarfs share similar formation
histories. 

\end{abstract}


\section{Introduction}

The subject of the formation of substellar objects has garnered significant
attention in the past few years. It remains unclear whether brown dwarfs 
form from the collapse of very low mass cloud cores, such as produced by 
turbulent fragmentation \citep{pn02}, or whether they are created via ejection 
of low-mass embryos from the main reservoirs of gas in a molecular 
cloud \citep{rc01,bat02}. These two possibilities can be investigated 
by comparing the characteristics of newborn substellar objects to those 
of their T Tauri counterparts at stellar masses.

The paradigm for low-mass star formation begins with the collapse
of a molecular cloud core onto a protostar and circumstellar accretion
disk \citep{als87}. The protostar continues to grow
via magnetospheric accretion, whereby the inner regions of
the circumstellar disk are disrupted by the stellar magnetosphere,
which then channels viscously accreting material out of the disk plane and
towards the star along magnetic field lines \citep{kon91,shu94}.
Signposts of this process, which have been directly observed in
classical T Tauri stars (hereafter, CTTSs), include:
infrared (IR) emission from dust at a range of temperatures in the disk
heated by stellar irradiation and viscous dissipation 
(e.g., \citet{mey97,muz03a}); blue/UV continuum excess
emission from the accretion shock formed as accreting material falls
onto the stellar surface \citep{har95,val93,gul98,cg98};
broad permitted emission lines produced in the ballistic magnetospheric gas 
flows (\citet{muz01}, hereafter MCH); forbidden emission lines produced in 
accretion-powered winds and jets \citep{har95}.

Recent studies have begun to examine these various signposts 
near and below the hydrogen burning limit to ascertain whether the same 
formation mechanism applies at both stellar and substellar masses.
High-resolution optical 
spectroscopy of young brown dwarfs has indeed provided considerable evidence 
for magnetically-mediated disk accretion, including broad permitted emission 
line profiles (especially \ha) and optical continuum veiling
(\citet{wb03}, hereafter, WB03; \citet{muz03b}, hereafter, M03; \citet{jay03b}).
Furthermore, detection of IR excess emission in many objects confirms 
the presence of irradiated circumstellar disks 
\citep{com98,com00,luh99,mue01,nat02,jay03a}.

In this paper, we present high-resolution spectroscopy of \ha emission line 
profiles for young low-mass objects, identify likely accretion candidates,
and estimate mass accretion rates through profile modeling. 
This work extends measurements of accretion to the lowest masses yet 
examined, approaching the planetary mass regime ($M/\msun\sim0.015$), and 
significantly increases the number of substellar objects with accretion 
measurements.  When these results are combined with our previous studies of 
accretion at higher masses, we are able to trace the change in accretion rates 
from stars to brown dwarfs across two orders of magnitude in mass. 

\section{Observations}

For this study, we considered spectroscopically confirmed
low-mass members of the nearest star forming regions that
are optically accessible ($I\lesssim20$). Preference was given to objects
with strong H$\alpha$ emission. The resulting sample consists of 39 members
of the Taurus and Chamaeleon~I star-forming regions taken from
\citet{bri02}, \citet{luh03a}, Luhman (2004a,b,c),
and Luhman (in preparation), which are listed in Table~\ref{data}.  
The youth and membership of all of the objects in Taurus and
Chamaeleon I have been established in these previous studies using
low-resolution spectroscopy, determining spectral types from various
atomic and molecular features and selecting against main sequence dwarfs
or giants using gravity-sensitive K~I and Na~I absorption lines
that are indicative of pre-main-sequence sources. The membership of some
of these objects has been independently confirmed by the presence of
reddening in their spectra and their positions above the main sequence for
the distance of each region, which indicate that they cannot be field
dwarfs in the foreground or the background of the cloud, respectively.
The detections of accretion presented in this work represent additional
evidence of youth, since such strong broad H$\alpha$ emission is not seen
in field dwarfs.

\begin{figure}
\plotone{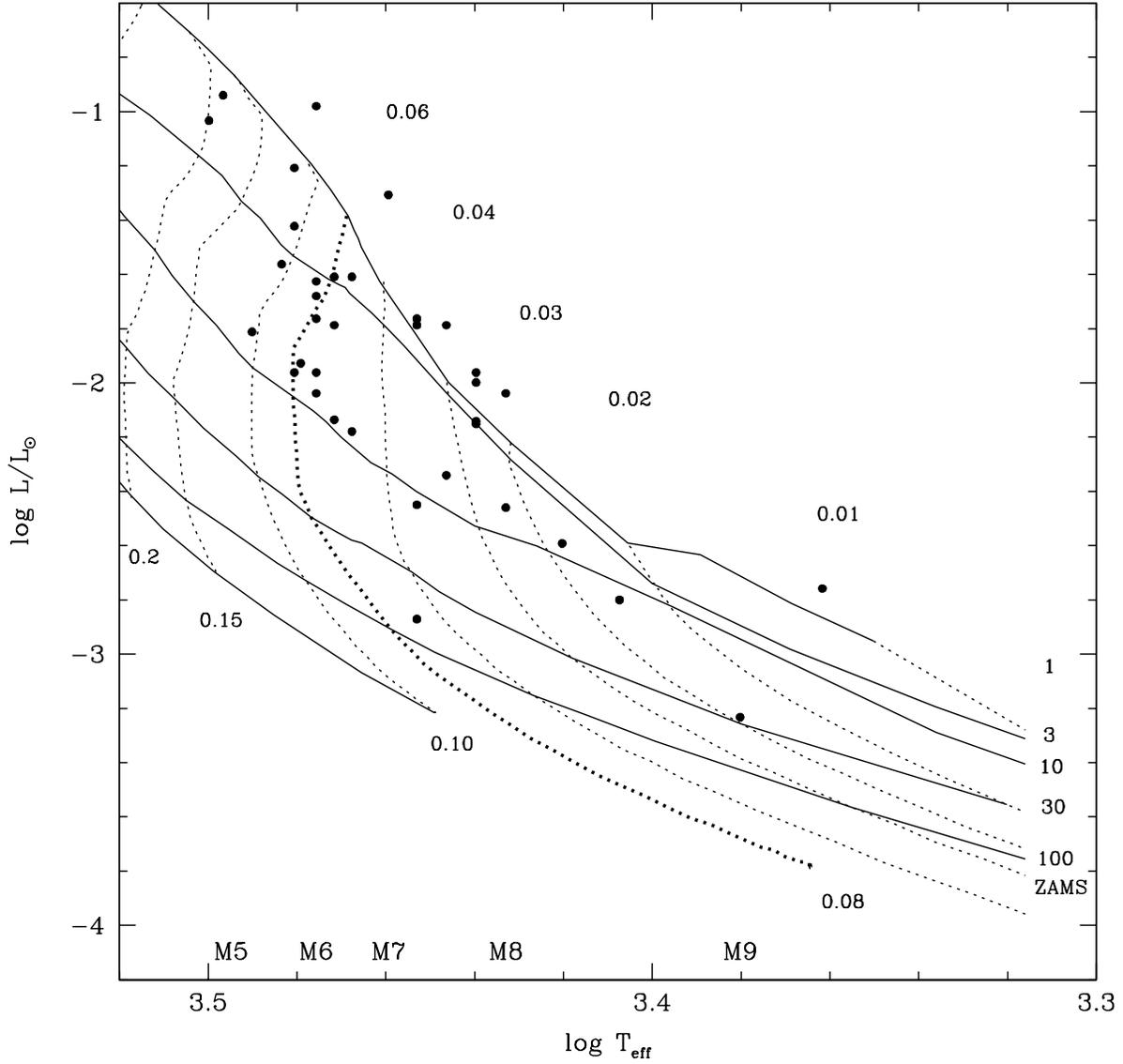}
\caption{H-R diagram for the targets of the high-resolution spectroscopy in
this work shown with the theoretical evolutionary models of
\citet{bar98} ($M/\msun>0.1$) and \citet{cha00} ($M/\msun\leq0.1$),
where the mass tracks ({\it dotted lines}) and isochrones ({\it solid lines})
are labeled in units of $\msun$ and Myr, respectively.
\label{hrd}}
\end{figure}

Our targets are placed on the Hertzsprung-Russell (H-R) diagram in
Figure~\ref{hrd} using temperatures and luminosities estimated in the
studies cited above. According to the evolutionary models of \citet{bar98}
and \citet{cha00}, $\sim25$ of these sources are brown dwarfs, one of which 
has a mass of only $\sim0.01$~$\msun$.
On the nights of 2003 December 30 through 2004 January 2, we obtained
spectra of these targets with the Magellan Inamori Kyocera Echelle (MIKE)
on the Magellan~II 6.5~m telescope at Las Campanas Observatory.
The instrument provided full coverage from 3200 to 10000~\AA\ at a
resolution near 20,000 with a $1\arcsec$ slit.
The exposure times ranged from 300 to 3600~s.
 
The MIKE data were bias-subtracted and trimmed using standard
IRAF\footnote{IRAF is distributed by the National Optical Astronomy
Observatories, which are operated by the Association of Universities for
Research in Astronomy, Inc., under cooperative agreement with the National
Science Foundation.} routines. 
The processed frames were then cleaned of cosmic rays with the {\sl crutils}
package. Since there was no signal in the blue camera spectra and the first
few orders of the red camera spectra, we only extracted one dimensional spectra
for the 22 reddest orders, providing an effective wavelength coverage of
about 5700 to 10000~\AA.  The spectra were extracted with the {\sl apall}
routine in the IRAF {\sl echelle} package, and wavelength calibrated
using ThAr lamp spectra.

\section{Analysis}

\subsection{Observed \ha Profiles}

The observed \ha profiles for our entire sample are shown in
Figure~\ref{profiles}.  Profile shapes and strengths closely resemble
those from previous observations of young low-mass objects.
Emission is detected in all of the targets; $\sim60\%$ of them exhibit narrow
(FWHM$\lesssim200$~km~s$^{-1}$), symmetric profiles, while the remainder
display broader, asymmetric profiles, occasionally with superimposed
blueshifted or central absorption reversals.  The former type of
emission is likely due to chromospheric activity on the substellar
surface, while the latter is probably produced in accretion flows
channeled from a circumstellar disk by the substellar magnetic field 
(\citet{muz00}; M03; WB03).  At least three objects, 2M J04554757+3028077,
2M J04141188+2811535, and 2M J11013205-7718249, show blueshifted absorption
components superimposed on the accretion emission profiles, indicating
accretion-powered mass loss analogous to that seen in higher-mass
T Tauri stars.  One of these, 2M J11013205-7718249, shows an extreme P Cygni
profile, with no evident emission at blue velocities $\lesssim -50$ km s$^{-1}$,
and may be entirely formed in the wind.  Such wind-dominated profiles are
occasionally seen in higher-mass CTTSs with large mass loss rates,
such as DR Tau (e.g., MCH).  Since mass loss typically scales with
accretion (Hartigan et al. 1995), it is somewhat surprising to see this
behavior in substellar objects, given their generally small accretion rates.
Substellar objects of this type are thus worthy of further scrutiny.

\begin{figure}
\plotone{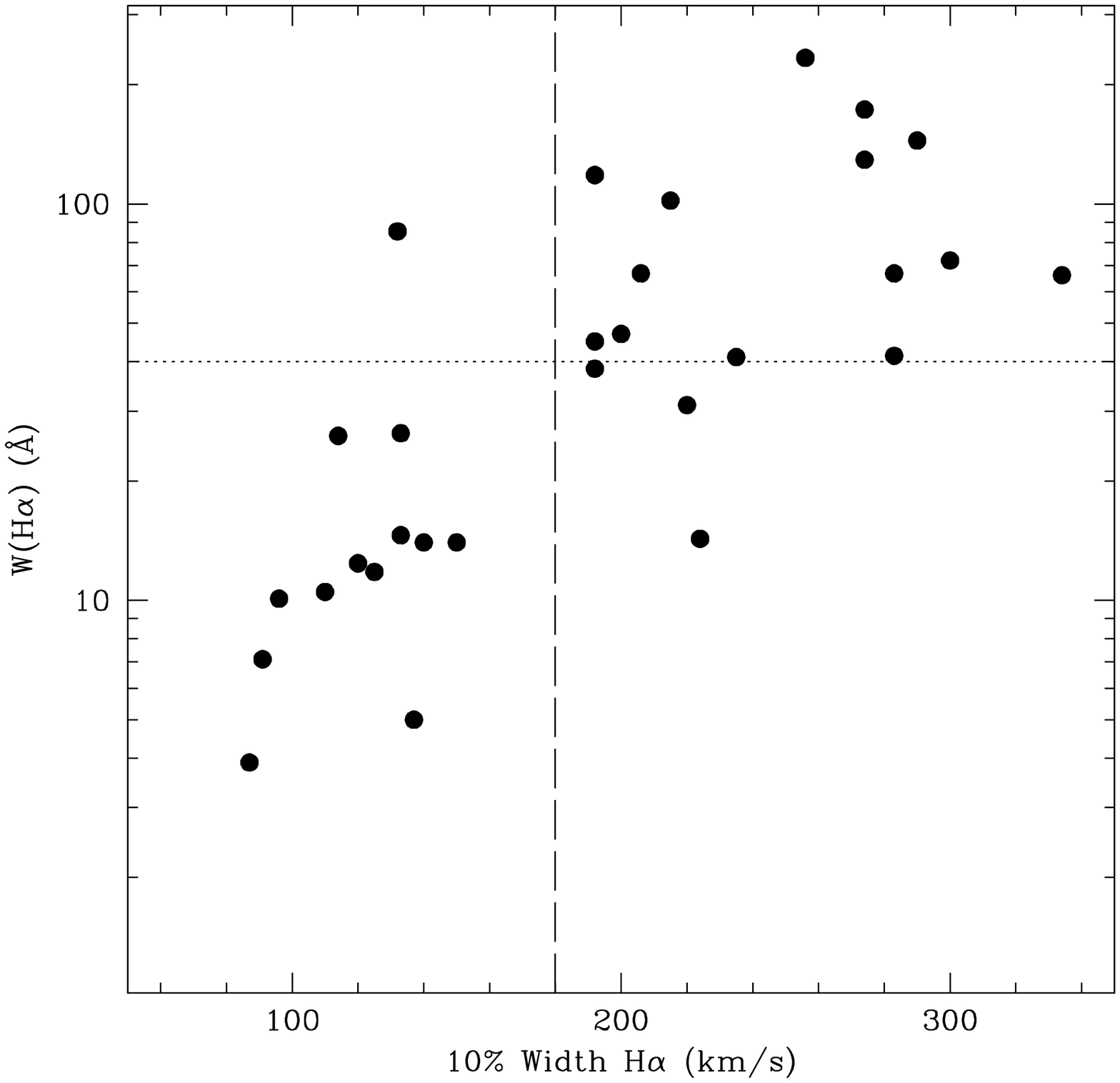}
\caption{
Equivalent widths of H$\alpha$ vs. the 10\% widths of H$\alpha$ for our
sample of young stars with spectral types M6 and later ($M/\msun\lesssim0.08$).
The horizontal dotted line indicates the W(H$\alpha$) = 40{\AA }
nominal limit between
weak-lined T Tauri stars and Classical T Tauri
stars for spectral types M6-7.5, as
defined by \citet{wb03}.
The vertical dashed line is our adopted
threshold separating accretors and chromospherically active stars.
\label{widths}}
\end{figure}

Based on a comparison between \ha emission line equivalent widths
($W_{\lambda}$), \ha full-widths at 10\% of the peak intensity ($V_{10}$), 
and spectral types, WB03
determined an empirical boundary between accretion and chromospheric activity
of $W_{\lambda}>40$~{\AA} and $V_{10}>270$ km s$^{-1}$ for young low-mass
objects with spectral types later than M6.
In Figure~\ref{widths}, we examine the same issue for our sample, 
including for the first time objects at $M/\msun\lesssim0.05$.
We see a significant number of objects later than M6 with 
$W_{\lambda}>40$~{\AA}, but with $V_{10}$ below the nominal
270~km~s$^{-1}$ threshold adopted by WB03.  This has also been noted
by \citet{jay03b}, who adopted a lower limit of $V_{10}>200$ km s$^{-1}$
based on observations of three accretors with spectral types later than M6.
The lower line widths are likely a reflection
of intrinsically smaller accretion flow velocity fields due to
the smaller object masses and radii; the maximum velocity from infall
is $V_{infall} = \sqrt{2GM_*/R_*}$, which yields an upper
limit to infall line widths of about 195 km s$^{-1}$ for the lowest-mass
objects in our sample.  
In order to define the accretor subsample, we have adopted a 10\% width
criterion $V_{10}>180$ km s$^{-1}$, which is at the upper end of
the most obvious break between \ha emission characteristics.
A total of 17 objects meet this criterion.

We may be missing some objects with broad accretion components
that are undetectable given the signal-to-noise of our data.
Using the accretion models described below, we predict a minimum
measurable mass accretion rate of $\sim 10^{-12} \; \msunyr$ for
objects with $M \sim 0.05 \; \msun$ (this limit will vary somewhat
with the mass of the central object because of the change in
the underlying photospheric continuum).  Thus, some of the objects
not identified as accretors in this work may in fact still be accreting,
but below this extremely low rate.  However, given the significant separation
of two populations apparent in the 10\% width distribution, as opposed to
a continuous distribution, we believe the number of missing accretors
is likely to be small.

\subsection{Model \ha Profiles}

Our goal in this paper is to determine mass accretion rates for our
substellar sample.  There are only two primary indicators currently
available for this purpose: measuring veiling continuum excess or
modeling \ha emission profiles.  For objects near or below the substellar
limit, the former method has been applied in just a few cases by WB03 and M03.
Unfortunately, veiling is unmeasurable for the large majority of very low
mass objects, given their generally tiny accretion rates (M03).
Thus, modeling of \ha is the most widely applicable primary diagnostic,
and we use this method here.
 
Previously, we have developed radiative transfer models of
permitted emission line profiles for both solar-type CTTSs
(MCH) and their low-mass stellar and substellar counterparts
(M03). These models treat
line emission produced by magnetospheric accretion flows,
where gas accreting through a circumstellar disk is channeled
onto the star by the stellar magnetic field.  The model
assumes an ideal dipolar field geometry, and a specified
gas temperature distribution and mass accretion rate;
the gas density distribution is determined directly from
the geometry and the gas velocity as a function of the mass and radius of 
the star. We defer most of the theoretical details to MCH and M03.

 
In brief, for this paper we have adopted the same accretion model,
using several values of the stellar mass (0.025 - 0.15~$\msun$)
and radius (0.25 - 1~$\rsun$) that are representative of our sample
(see Table~\ref{data}).  The actual mass range will change
somewhat depending on which evolutionary tracks are used; however,
this should not affect our modeling applications since the model
gas velocities, and, hence the line widths, are only sensitive to
$\sqrt{M}$, and the derived masses are unlikely to be off by more
than a factor of two.  We employ the same temperature
constraints as those found to reproduce various emission profiles
and line ratios for CTTSs
with well-determined accretion rates from other observations.
As we discussed in M03, this assumption should be
adequate since the gas is probably heated mechanically through processes
independent of spectral type.  The temperature constraints
result in an inverse
proportionality between the gas temperature and density.
Due to the lack of opacity broadening in any of the observed \ha
profiles, we can rule out $\mdot \gtrsim 10^{-9} \; \msunyr$ (M03).
For lower values of $\mdot$, the gas temperature can be fairly well-
constrained; for $T \gtrsim 10,000$ K, the gas is already
completely ionized, so the line flux will not change appreciably,
while below this level, there is not enough line optical depth
to produce emission at observed levels.
Further details on the parameter constraints are found in M03;
here, we apply the same methodology.
 

We consider the likely accretors identified in the previous section
that have spectral types M6 and later.  The wind-dominated object
2M J11013205-7718249 lacks emission in the blue wing, thus we cannot
reliably match an accretion model and do not include
it in the following analysis.  
For each of the remaining accretor \ha profiles, we constructed a model
by adopting the magnetospheric temperature
and size constraints discussed in M03,
using the model substellar mass and radius closest to the empirically
determined values, and finally varying only the mass accretion
rate and inclination angle to find the best match to the observed
profile shape and flux.  A range of accretion rates was explored,
$10^{-12} < \mdot < 10^{-10} \; \msunyr$.  The lower limit
corresponds to the level where accretion emission is produced at
an undetectable level, while values above the upper limit would
produce too much flux both in the wings (from opacity broadening)
and in the line overall compared to the data in our sample.
The inclination angle was varied from pole-on to edge-on
orientations of the star/disk system.  The comparisons were made
by eye; in the cases where the profile exhibited central or
blueshifted absorption components, only the high-velocity line wings
unaffected by absorption were considered.

\begin{figure}
\plotone{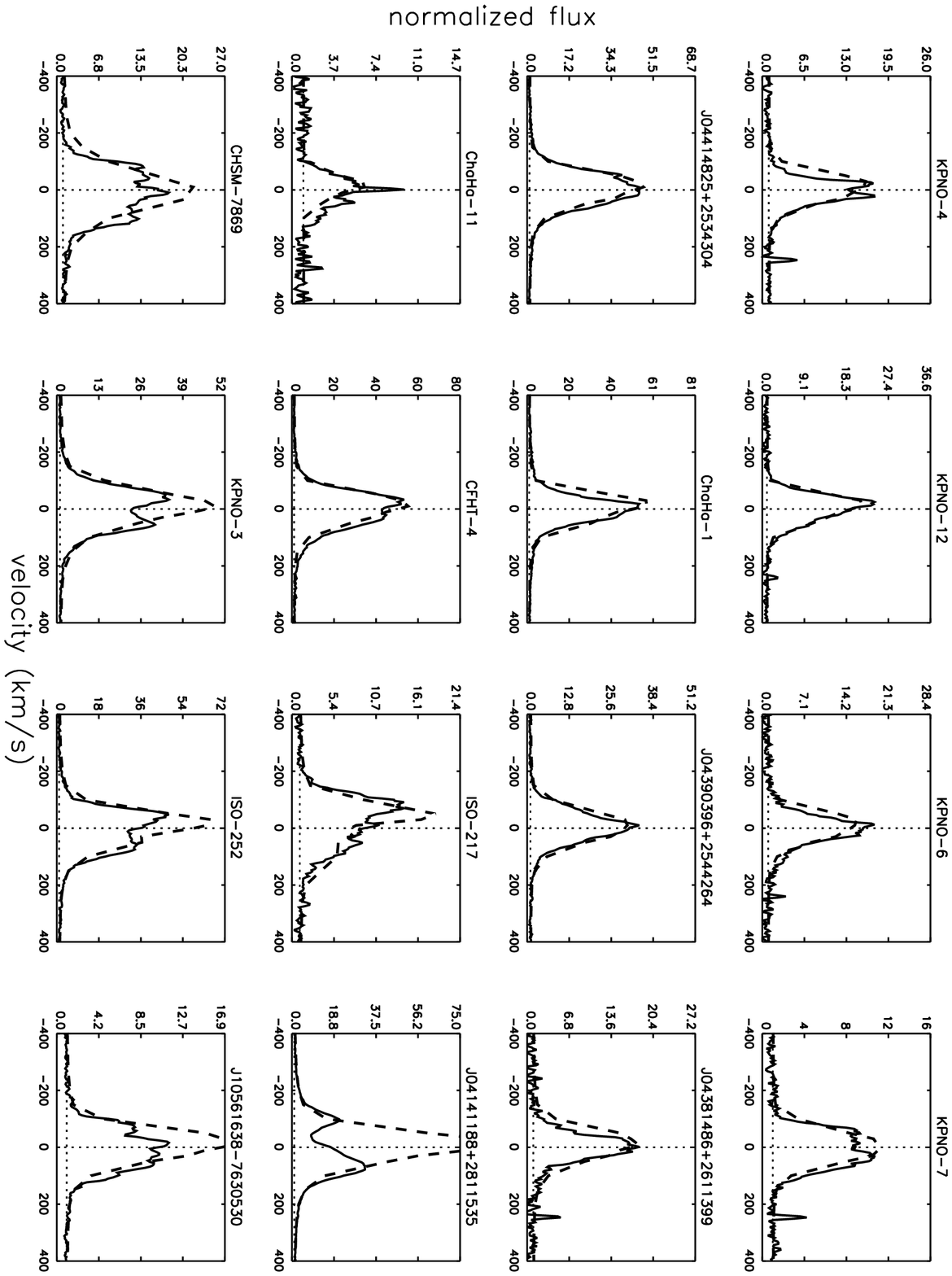}
\caption{Comparisons of observed ({\it solid}) and model ({\it dashed}) 
\ha emission line profiles for objects in our sample
that have profiles indicative of accretion (Table~\ref{model_param}).
\label{models}}
\end{figure}

Our best model matches are shown in Figure~\ref{models},
with parameters listed in Table~\ref{model_param}.
The accretion rates for these objects and the limits for the remainder of 
our sample are also provided in Table~\ref{data}.
We do not include the results for KPNO 4 as the best model match does
not agree particularly well in terms of profile shape -- the observed
profile is quite symmetric, with a central absorption reversal strongly
suggestive of chromospheric emission.  The $V_{10}$ width and $EW$ for
this object are marginal for accretion, and could also be explained by
chromospheric emission broadened by rapid rotation; unfortunately,
our spectrum does not have enough $S/N$ to measure $v{\sin}i$.

As mentioned in M03, there is a systematic uncertainty in our
model-derived $\mdot$ values of about a factor of 3-5.
This is mainly due to uncertainties in the size of the accretion flow,
which is not well-constrained for our particular
sample.  However, observations of infrared excesses around brown dwarfs
in general infer inner disk hole sizes of roughly $2-5 \; R_*$,
similar to our adopted value (Natta et al. 2002; Liu et al. 2003;
Mohanty et al. 2004).

\section{Discussion}

\begin{figure}
\plotone{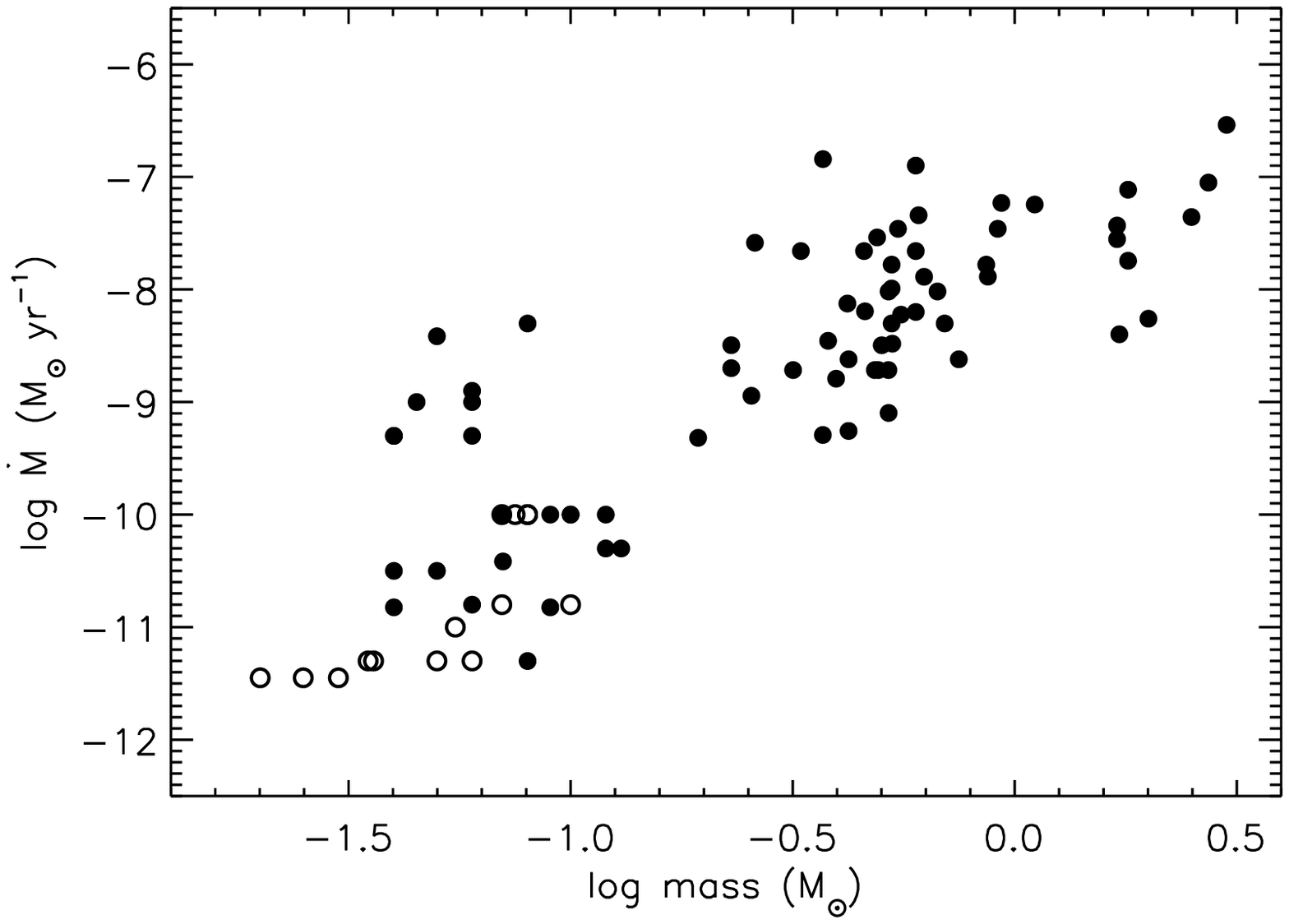}
\caption{Mass accretion rate as a function of mass. Filled circles are from 
\citet{gul98}, \citet{wg01}, WB03, M03, \citet{cal04}, and Natta et al. (2004),
 and open circles are from this work.
\label{mass_mdot}}
\end{figure}

Our results more than triple the number of known substellar accretors
(those with spectral types later than M6) to 16, and extend
measurements of mass accretion rates down to the lowest masses yet identified.
The new accretion rates for our sample of low-mass stars
and brown dwarfs, combined with previous estimates for other brown dwarfs
and more massive stars, are plotted as a function of mass
in Figure~\ref{mass_mdot}.
The accretion rates for substellar objects are extremely small, mostly
$< 10^{-10} \; \msunyr$, several of which are among the lowest values
measured to date at $\mdot = 4-5 \times 10^{-12} \; \msunyr$.
These data continue a trend that we have noted previously for the more
massive objects in which mass accretion rates depend steeply on mass;
here, we find a correlation of $\mdot \propto M^{2.1}$.
In addition, the change in accretion rate with mass is fairly continuous 
across six orders of magnitude in $\mdot$ and more than two orders of 
magnitude in mass, which suggests that the same formation mechanism is
responsible for both the low- and high-mass objects.
Folding in previous work mentioned above,
and our own studies of intermediate mass young stars (Calvet et al.\ 2004),
there is now significant evidence
that the disk accretion paradigm applies to a range of over two orders
of magnitude in mass, from $0.02 - 3 \; \msun$.
Thus, free-floating stellar and substellar objects in this mass range,
comprising all but the most massive O and B stars, likely form from the
collapse of a molecular cloud core onto an accretion disk and central object.

There is considerable scatter in $\mdot$ for a given mass all across
the mass spectrum, as has been noted previously for the higher-mass objects.
Most of this scatter probably reflects intrinsic differences from
source to source, and may be due to age differences, since viscous evolution
expects a general decrease in the accretion rate with time (Hartmann et al.
1998), or differences in initial conditions.  We note that the accretion rate
measurements for low mass members of Ophiuchus from Natta et al. (2004)
appear to be larger on average by about a factor of 5-10 than those in
Taurus, Chamaeleon I, and IC 348.  This may be an indication
of the slightly younger age of Ophiuchus, or possibly different initial
conditions such as the larger envelope infall rates that characterize
protostellar objects in that region.


Our results also imply that objects down to masses of 0.02~$\msun$ harbor
significant, ordered magnetic fields.  Accretion infall cannot occur
unless there is a magnetic field of sufficient strength to truncate
the inner disk and channel material onto the surface of the substellar object.  
Given the model-derived mass accretion rates, inner disk hole size implied
by our adopted inner magnetospheric radius, and empirical
mass and radius estimates, we predict magnetic field strengths on
the order of 100-200 G for the lowest-mass accretors in our sample,
using the theoretical relations given by K\"onigl (1991).
Significant magnetic fields are also likely to exist on the nonaccretors,
given their signatures of chromospheric activity similar to that seen in
M and L field dwarfs (e.g. Mohanty \& Basri 2003).

Finally, given the object masses and typical ages of $\sim1$~Myr,
the mass accretion rates in Figure~\ref{mass_mdot} will have little or no 
effect on the remaining history of the young stars and brown dwarfs we have
considered. Presumably, most of the mass is accreted in the earliest phases
of evolution, either in the initial collapse of the cloud core 
or via outbursts initiated by disk instabilities.
As such, it will be important to eventually identify and study
embedded protostellar counterparts at the lowest masses.






\acknowledgements

We thank the staff at Las Campanas Observatory for their support
of these observations. K. L. was supported by grant NAG5-11627 from the
NASA Long-Term Space Astrophysics program.
2MASS is a joint project of the University of Massachusetts
and the Infrared Processing and Analysis Center/California Institute
of Technology, funded by the National Aeronautics and Space
Administration and the National Science Foundation.

\begin{deluxetable}{lllllllllll}
\setlength{\tabcolsep}{0.04in}
\tabletypesize{\tiny}
\tablewidth{0pt}
\tablecaption{Data for Observed Sample \label{data}}
\tablehead{
\colhead{} &
\colhead{} &
\colhead{} &
\colhead{} &
\colhead{$T_{\rm eff}$\tablenotemark{b}} &
\colhead{$L_{\rm bol}$\tablenotemark{b}} &
\colhead{Radius} &
\colhead{Mass} &
\colhead{$W_{\lambda}$(H$\alpha$)} &
\colhead{H$\alpha$ 10\% Width} &
\colhead{log \.M} \\
\colhead{ID} &
\colhead{$\alpha$(J2000)\tablenotemark{a}} &
\colhead{$\delta$(J2000)\tablenotemark{a}} &
\colhead{Spectral Type\tablenotemark{b}} &
\colhead{(K)} &
\colhead{($\lsun$)} &
\colhead{($\rsun$)} &
\colhead{($\msun$)} &
\colhead{(\AA)} &
\colhead{(km s$^{-1}$)} &
\colhead{($\msun$~yr$^{-1}$)}}
\startdata
         KPNO~4 &   04 27 28.00 & $+$26 12 05.3 &   M9.5 &   2300 & 0.0023 &   0.30 &  0.011 &  -38.4 &    192 &  $<$-12 \\
        KPNO~12 &   04 19 01.27 & $+$28 02 48.7 &     M9 &   2400 & 0.00082 &   0.17 &  0.020 &  -66.8 &    206 &  -11.4 \\
         KPNO~6 &   04 30 07.24 & $+$26 08 20.8 &   M8.5 &   2555 & 0.0021 &   0.23 &  0.025 &  -41.1 &    235 & -11.4 \\
         KPNO~7 &   04 30 57.19 & $+$25 56 39.5 &  M8.25 &   2632 & 0.0033 &   0.28 &  0.030 &  -31.1 &    220 & -11.4 \\
 2M~J11013205$-$7718249 &   11 01 32.05 & $-$77 18 25.0 &     M8 &   2710 &  0.0044 &   0.30 &  0.035 &  -66.8 &    283 & \nodata \\
     CHSM~17173 &   11 17 23.83 & $-$76 24 08.6 &     M8 &   2710 &  0.011 &   0.48 &  0.030 & -85.3 &  132  & $<$-12 \\
 2M~J04414825$+$2534304 &   04 41 48.25 & $+$25 34 30.5 &  M7.75 &   2752 & 0.0086 &   0.41 &  0.035 & -233.7 &  256 & -11.3 \\
 Cha~H$\alpha$1 &   11 07 16.69 & $-$77 35 53.3 &  M7.75 &   2752 &  0.013 &   0.50 &  0.035 & -118.3 &    192 & -11.3 \\
 Cha~H$\alpha$7 &   11 07 37.76 & $-$77 35 30.8 &  M7.75 &   2752 &  0.012 &   0.48 &  0.035 &  -26.0 &    114 & $<$-12 \\
         CFHT~3 &   04 36 38.94 & $+$22 58 11.9 &  M7.75 &   2752 & 0.0088 &   0.41 &  0.035 &  -10.5 &    110 & $<$-12 \\
         KPNO~2 &   04 18 51.16 & $+$28 14 33.2 &   M7.5 &   2795 & 0.0057 &   0.32 &  0.045 &   -3.9 &     87 & $<$-12 \\
         KPNO~5 &   04 29 45.68 & $+$26 30 46.8 &   M7.5 &   2795 &  0.019 &   0.59 &  0.040 &  -26.4 &    133 & $<$-12 \\
 2M~J04390396$+$2544264 &   04 39 03.96 & $+$25 44 26.4 &  M7.25 &   2838 &  0.019 &   0.57 &  0.050 & -102.0 &    215 & -11.3 \\
 2M~J04381486$+$2611399 &   04 38 14.86 & $+$26 11 39.9 &  M7.25 &   2838 & 0.0018 &   0.18 &  0.070 &  -47.0 &    200 & -10.8 \\
Cha~H$\alpha$11 &   11 08 29.27 & $-$77 39 19.8 &  M7.25 &   2838 & 0.0045 &   0.28 &  0.055 &  -14.3 &    224 &  -11 \\
 2M~J11011926$-$7732383 &   11 01 19.27 & $-$77 32 38.3 &  M7.25 &   2838 &  0.020 &   0.59 &  0.050 &  -10.1 &     96 & $<$-12 \\
         CFHT~4 &   04 36 38.94 & $+$22 58 11.9 &     M7 &   2880 &  0.054 &    0.94 &  0.060 & -129.3 &    274 & -11.3 \\
Cha~H$\alpha$12 &   11 06 38.00 & $-$77 43 09.1 &   M6.5 &   2935 &  0.028 &   0.65 &  0.075 &  -14.6 &    133 & $<$-12 \\
        ISO~138 &   11 08 18.51 & $-$77 30 40.8 &   M6.5 &   2935 & 0.0081 &   0.35 &  0.065 &  -12.4 &    120 & $<$-12 \\
        ISO~217 &   11 09 52.16 & $-$76 39 12.8 &  M6.25 &   2962 &  0.028 &   0.64 &  0.080 &  -66.1 &    334 & -10 \\
Cha~H$\alpha$10 &   11 08 24.04 & $-$77 39 30.0 &  M6.25 &   2962 & 0.0089 &   0.36 &  0.070 &   -7.1 &     91 & $<$-12 \\
 2M~J04141188$+$2811535 &   04 14 11.88 & $+$28 11 53.5 &  M6.25 &   2962 &  0.019 &   0.53 &  0.075 & -250.0 & 362 & -10 \\
      CHSM~7869 &   11 06 32.77 & $-$76 25 21.1 &     M6 &   2990 &  0.028 &   0.39 &  0.070 &  -72.0 &    300 & -10 \\
        KPNO~14 &   04 33 07.81 & $+$26 16 06.6 &     M6 &   2990 &   0.11 &    1.2 &   0.10 &  -14.0 &    140 & $<$-12 \\
         KPNO~3 &   04 26 29.39 & $+$26 24 13.8 &     M6 &   2990 &  0.020 &   0.53 &  0.080 & -144.6 &    290 & -10 \\
        ISO~252 &   11 10 41.42 & $-$77 20 48.1 &     M6 &   2990 &  0.027 &   0.62 &  0.080 & -173.1 &    274 & -10 \\
 2M~J11070324$-$7610565 &   11 07 03.24 & $-$76 10 56.6 &     M6 &   2990 &  0.013 &   0.43 &  0.070 &   -9.3 & 169 & $<$-12 \\
2M~J11080234$-$7640343 &   11 08 02.35 & $-$76 40 34.4 &     M6 &   2990 &  0.024 &   0.58 &  0.080 &   -5.0 &    120 & $<$-12 \\
 2M~J11075993$-$7715317 &   11 07 59.93 & $-$77 15 31.8 &  M5.75 &   3024 &  0.067 &    0.95 &  0.12 &   -8.4 &    178 & $<$-12 \\
 2M~J11072443$-$7743489 &   11 07 24.44 & $-$77 43 49.0 &  M5.75 &   3024 &  0.042 &   0.75 &  0.11 &   -8.8 &    114 & $<$-12 \\
 2M~J11173792$-$7646193 &   11 17 37.93 & $-$76 46 19.4 &  M5.75 &   3024 &  0.013 &   0.42 &  0.080 &  -11.8 &    125 & $<$-12 \\
 2M~J04554801$+$3028050 &   04 55 48.01 & $+$30 28 05.0 &   M5.6 &   3014 &  0.014 &   0.44 &  0.075 &  -14.0 &    150 & $<$-12 \\
 2M~J10561638$-$7630530 &   10 56 16.38 & $-$76 30 53.0 &   M5.6 &   3014 &  0.031 &   0.64 &  0.10 &  -41.4 &    283 & -10.8 \\
 2M~J10580597$-$7711501 &   10 58 05.98 & $-$77 11 50.1 &  M5.25 &   3091 &  0.018 &   0.47 &  0.10 &   -5.0 &    137 & $<$-12 \\
 2M~J04554757$+$3028077 &   04 55 47.57 & $+$30 28 07.7 &  M4.75 &   3161 &  0.098 &    1.0 &   0.20 &  -24.5 &  320 & \nodata \\
            T23 &   11 06 59.07 & $-$77 18 53.6 &  M4.25 &   3138 &   0.12 &    1.2 &   0.18 &    -7.0 &    247 & \nodata \\
            T3W &   10 55 59.73 & $-$77 24 39.9 &   M3.5 &   3850 &   0.27 &    1.2 &   0.73 &   -9.5 &    259 & \nodata \\
      CHXR49~NE &   11 11 54.00 & $-$76 19 31.1 &   M2.5 &   3430 &   0.34 &    1.7 &   0.44 &      0 &      0 & \nodata \\
            T3E &   10 55 59.73 & $-$77 24 39.9 &   M0.5 &   3342 & \nodata & \nodata &   0.35 &   -8.9 &    247 & \nodata \\
\enddata
\tablecomments{Units of right ascension are hours, minutes, and seconds, and
units of declination are degrees, arcminutes, and arcseconds.}
\tablenotetext{a}{2MASS Point Source Catalog.}
\tablenotetext{b}{\citet{bri02}, \citet{luh03a}, \citet{luh04a}, \citet{luh04b},
\citet{luh04c}, and Luhman (in preparation).}
\end{deluxetable}

\begin{deluxetable}{llllll}
\tablecaption{\ha Model Parameters for Accretor Subsample \label{model_param}}
\tablehead{
\colhead{ID} & \colhead{M$_*$ ($\msun$)} & \colhead{R$_*$ ($\rsun$)} &
\colhead{T$_{eff}$ (K)} & \colhead{i ($^{\circ}$)} &
\colhead{log $\mdot$ ($\msunyr$)}}
\startdata
KPNO~12 & 0.025 & 0.25 & 2600 & 45 & -11.4 \\
KPNO~6 & 0.025 & 0.25 & 2600 & 60 & -11.4  \\
KPNO~7 & 0.025 & 0.25 & 2600 & 75 & -11.4 \\
2M J04414825+2534304 & 0.025 & 0.25 & 2600 & 60 & -11.3 \\
Cha~{\ha}1 & 0.025 & 0.25 & 2600 & 45 & -11.3 \\
2M J04390396+2544264 & 0.025 & 0.25 & 2600 & 70 & -11.3 \\
2M J04381486+2611399 & 0.05 & 0.5 & 3000 & 60 & -10.8 \\
Cha~{\ha}11 & 0.05 & 0.5 & 3000 & 30 & -11 \\
CFHT~4 & 0.025 & 0.25 & 2600 & 55 & -11.3 \\
ISO~217 & 0.05 & 0.5 & 3000 & 65 & -10$^a$ \\
2M J04141188+2811535 & 0.05 & 0.5 & 3000 & 55 & -10 \\
CHSM~7869 & 0.15 & 1.0 & 3000 & 85 & -10 \\
KPNO~3 & 0.15 & 1.0 & 3000 & 60 & -10 \\
ISO~252 & 0.15 & 1.0 & 3000 & 45 & -10 \\
2M J10561638-7630530 & 0.05 & 0.5 & 3000 & 60 & -10.8 \\
\enddata
\tablecomments{All models calculated with magnetospheric radii $R_{mag} = 2.2-3$
$\rsun$ and maximum temperature $T_{max} = 12,000$ K.}
\tablenotetext{a}{Calculated with modified temperature distribution; see text.}
\end{deluxetable}

\begin{figure}
\plotone{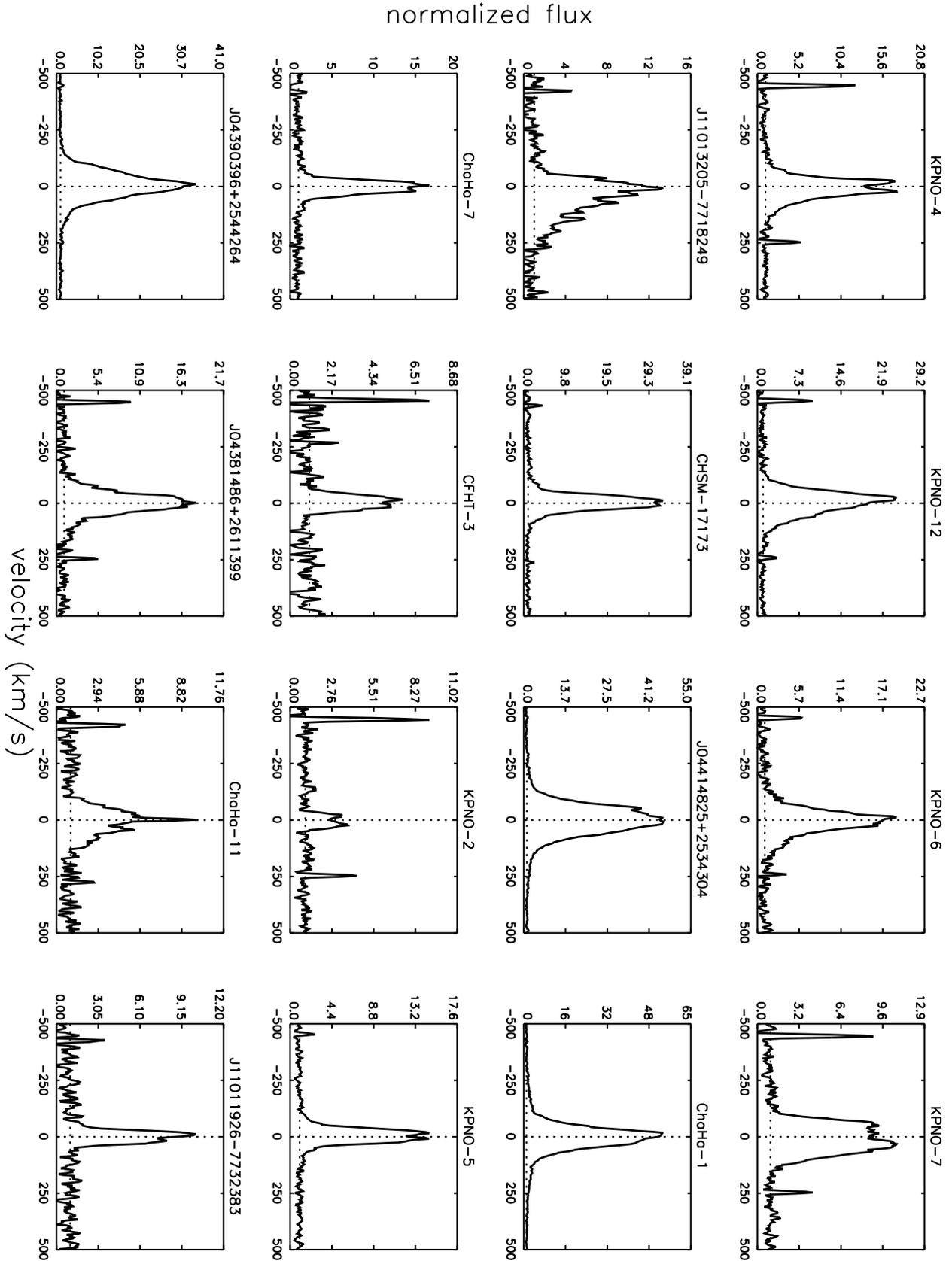}
\caption{Observed \ha emission line profiles. The spikes close to -500 and
+250 km/s seen in some spectra (e.g. J04381486+2611399 and KPNO-2)
are sky lines that could not be completely subtracted out.
\label{profiles}}
\end{figure}

\begin{figure}
\plotone{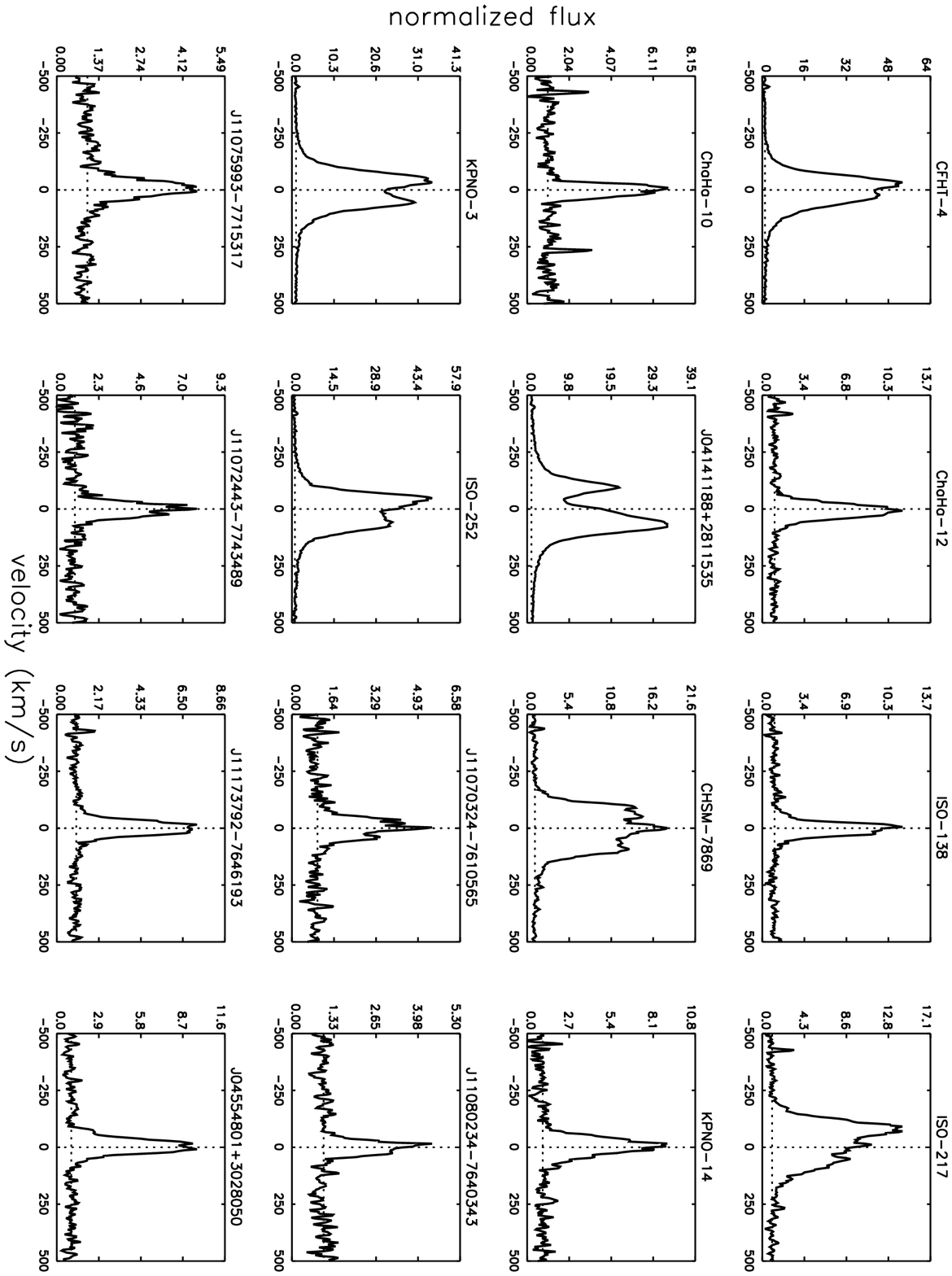}
\addtocounter{figure}{-1}
\caption{Continued.}
\end{figure}

\begin{figure}
\plotone{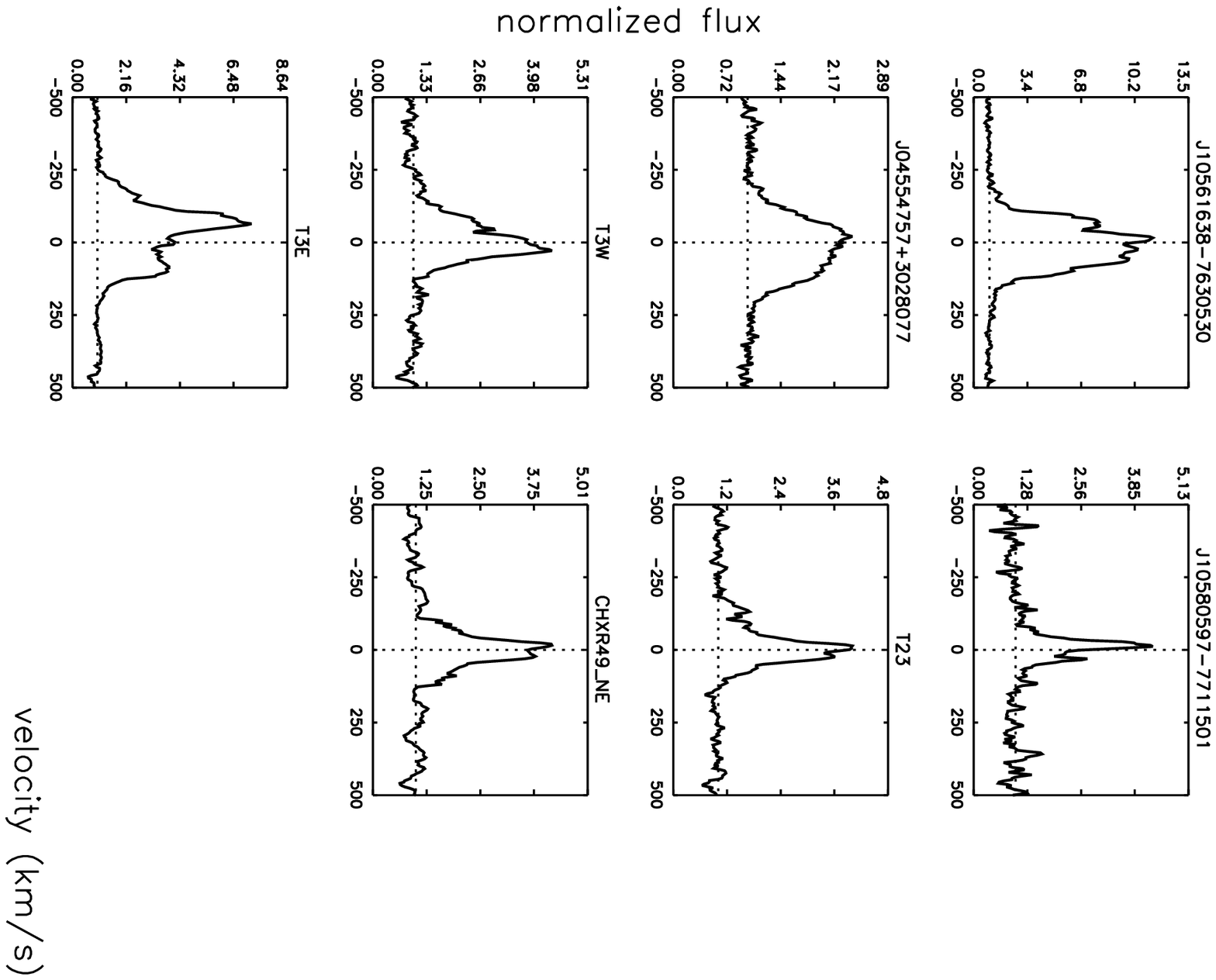}
\addtocounter{figure}{-1}
\caption{Continued.}
\end{figure}

\end{document}